\documentclass[11pt]{elsarticle}

\usepackage{xcolor}  
\usepackage{nicefrac}  
\usepackage{comment}  
\usepackage{graphicx}  
\usepackage{longtable}  
\usepackage{algorithmic,bm,latexsym,amsmath,amssymb}
\usepackage[ruled]{algorithm2e}

\newtheorem{bolddefinition}{\textbf{Definition}}

\DeclareMathOperator*{\argmax}{arg\,max}

\newcommand{\PT}{\mathit{PT}}
\newcommand{\opt}{\mathit{opt}}
\newcommand{\CS}{\mathit{CS}}

\newcommand{\UB}{\mathit{UB}}
\newcommand{\TSP}{\mathit{TSP}}
\newcommand{\DTSP}{\mathit{D\textnormal{-}TSP}}
\newcommand{\CFSS}{\mathit{CFSS}}
\newcommand{\DyPE}{\mathit{DyPE}}

\newcommand{\ConnectedSubsets}{\mathit{ConnectedSubsets}}

\newcommand{\search}{\mathbf{search}}

\newcommand{\mcP}{\mathcal{P}}
\usepackage[ruled]{algorithm2e}
\newcommand{\Comment}[1]{{\scriptsize{\tcp*[h]{#1}}}}  

\journal{Artificial Intelligence}

\begin{document}

\begin{frontmatter}

\title{Coalition Structure Generation on Graphs}

\author[l2]{Talal Rahwan}
\author[l3]{Tomasz Michalak}
\address[l2]{Masdar Institute of Science and Technology, UAE}
\address[l3]{University of Oxford, UK, and University of Warsaw, Poland}

\begin{abstract}
Two fundamental algorithm-design paradigms are \emph{Tree Search} and \emph{Dynamic Programming}. The techniques used therein have been shown to complement one another when solving the complete set partitioning problem, also known as the \emph{coalition structure generation} problem \cite{Rahwan:etal:14}. Inspired by this observation, we develop in this paper an algorithm to solve the coalition structure generation problem on graphs, where the goal is to identifying an optimal partition of a graph into connected subgraphs. More specifically, we develop a new depth-first search algorithm, and combine it with an existing dynamic programming algorithm due to \citet{Vinyals:etal:13}. The resulting hybrid algorithm is empirically shown to significantly outperform both its constituent parts when the subset-evaluation function happens to have certain intuitive properties.
\end{abstract}

\end{frontmatter}


\section{Introduction}

\noindent The \emph{coalition structure generation} problem is a fundamental problem in multi-agent systems research. It involves partitioning the set of agents into mutually disjoint coalitions so that the total reward from the resulting coalitions is maximized. Most of the literature on this topic assumes that the agents can split into teams (or \emph{coalitions}) in any way they like \cite{Rahwan:etal:15:AIJ}. In practice, however, some coalition structures may be inadmissible due to various constraints present in the problem domain. 

This paper considers one such class of problems, known as \emph{graph-restricted games} \cite{Myerson:1977}. Here, we are given a graph in which every node represents an agent, and every edge can be interpreted as a communication channel, or a trust relationship, which facilitates the cooperation between its two ends. A coalition is then feasible if and only if it induces a connected subgraph of $G$. The intuition here is that any two agents cannot belong to the same coalition unless they are able to communicate with one another, either directly through an edge, or indirectly through (some of) the other members of the coalition who collectively form a path between the two agents. Following convention, we will assume that $G$ is connected.\footnote{\footnotesize If $G$ is not connected, the coalition structure generation problem can be decomposed into smaller independent sub-problems, each having a connected graph.}

To the best of our knowledge, the two state-of-the-art algorithms for solving the coalition structure generation problem in graph-restricted games are: (i) a tree-search algorithm called $\CFSS$ \cite{Bistaffa:etal:14}, and (ii) a dynamic-programming algorithm called $\DyPE$ \cite{Vinyals:etal:13}. Each algorithm has its relative strengths and weakness compared to the other. In particular,
\begin{itemize}
\item $\CFSS$ is superior in that it is an anytime algorithm---its solution quality improves monotonically as computation time increases. As such, it can return a valid solution even if it was unable to run to completion, e.g., due to a failure or due to time constraints. $\DyPE$, on the other hand, is not an anytime algorithm, and so does not return interim solutions. Another advantage of $\CFSS$ is that it applies a branch-and-bound technique, which enables it to exploit the specifics of any given problem instance, resulting in (possibly significant) speedups. This is not possible with $\DyPE$ due to the absence of any such a branch-and-bound technique. Finally, $\CFSS$ uses very little memory compared to $\DyPE$; the latter requires storing in memory the solutions of different sub-problems, the number of which may be exponential (depending on the density of the graph).
\item On the other hand, $\DyPE$ is superior in terms of the computational complexity. For instance, given a complete graph of $n$ agents, $\DyPE$ runs in $O(3^n)$, while $\CFSS$ runs in $O(n^n)$. This is because the latter depends heavily on the branch-and-bound technique which, in the worst case, may fail to prune even a single solution, resulting in a brute-force search.
\end{itemize}

Since $\DyPE$ and $\CFSS$ have their own strengths and weaknesses relative to each other, it is desirable to develop an algorithm that has the best of both. Perhaps a promising direction is to combine $\DyPE$ with $\CFSS$, following the steps of \citet{Rahwan:etal:12}, who handled the general (not graph-restricted) coalition structure generation problem by combining a dynamic programming algorithm \cite{Rahwan:Jennings:08b} with a depth-first search algorithm \cite{Rahwan:09a}, thus obtaining the best of both. In our case, however, $\CFSS$ and $\DyPE$ are built on entirely different search-space representations (see Section~\ref{sec:relatedWork}), which makes it hard to combine the two algorithms elegantly and efficiently.

With this in mind, the contribution of this paper are as follows:

\begin{itemize}
\item We develop $\TSP$---a new depth-first search algorithm specifically designed to be compatible with  $\DyPE$.
\item We show how to modify both $\TSP$ and $\DyPE$ such that they complement one another when merged into a single hybrid algorithm, called $\DTSP$.
\item We empirically evaluate $\DTSP$ in randomly-generated super-subadditive settings, and show that it significantly outperforms its constituent parts.
\end{itemize}

The remainder of the paper is structured as follows. The main notation is introduced in Section~\ref{sec:preliminaries}. The existing dynamic-programming algorithm---$\DyPE$---and the existing tree-search algorithm---$\CFSS$---are described in Section~\ref{sec:relatedWork}. Our new tree-search algorithm---$\TSP$---is introduced in Section~\ref{sec:TSP}. The hybrid algorithm---$\DTSP$---is introduced in Section~\ref{sec:D-TSP}. 
Finally, Section~\ref{sec:conclusions} concludes the paper and discusses future directions.


\section{Preliminaries}\label{sec:preliminaries}

\noindent A graph-restricted game is a tuple, $(A,v,G)$, where $A$ is the set of agents, $v:2^A\to\mathbb{R}$ is a characteristic function that evaluates each coalition of agents, and $G=(A,E)$ is a graph whose set of nodes is $A$, and whose set of edges, $E$, specifies which agents are connected to each other. The number of agents in $A$ will be denoted by $n$.

For every coalition, $C\subseteq A$, let $\mcP^C$ denote the set of partitions of $C$, also known as coalition structures over $C$.\footnote{\footnotesize The terms ``partition'' and ``coalition structure'' will be used interchangeably throughout the paper, as common practice in the literature.} Given our focus on characteristic function games, the value of a partition is simply the sum of the values of the coalitions therein.\footnote{\footnotesize The coalition structure generation problem has also been studied in games where partitions are evaluated differently, e.g., due to the presence of externalities \cite{Michalak:09,Rahwan:etal:12:AIJ}. However, these are out of the scope of this paper, and are the focus of future work.} More formally, for every $C\subseteq A$, and every $P\in\mcP^C$, the value of partition $P$ is:
$$
V(P)=\sum_{p\in P}v(p).
$$
We will denote by $opt(C)$ an optimal partition of $C$, and by $v^*(C)$ the value of such a partition. More formally,
\begin{eqnarray}
\opt(C)\in\argmax_{P\in\mcP^C}V(P) &\ \ \ \textnormal{  and  } &\ \ \ v^*(C)=\max_{P\in\mcP^C}V(P).\nonumber
\end{eqnarray}
In a graph-restricted game, $(A,v,G)$, we say that a coalition, $C\subseteq A$, is \emph{connected in $G$} if and only if $C$ induces a connected subgraph of $G$. Moreover, for every $C\subseteq A$, we will denote by $\ConnectedSubsets(C,G)$ the set of all non-empty subsets of $C$ that are each connected in $G$. Similarly, we will denote by $\mcP^C_G$ the set of every partition in $\mcP^C$ whose coalitions are connected in $G$. More formally, $\mcP^C_G = \{P\in\mcP^C : P\cap \ConnectedSubsets(C,G) = P\}$. Then, given $(A,v,G)$, the coalition structure generation problem is to find an optimal partition of $A$, defined as follows:
$$
\CS^* \in \argmax_{P\in\mcP^A_G} V(P).
$$
Next, we define \emph{weakly super-subadditive games}. To this end, recall that a game $(A,v)$ is \emph{weakly superadditive} if: $v(C\cup C')\geq v(C)+v(C')$ for any two disjoint coalitions $C,C'$ (i.e., merging any two coalitions is never harmful). Conversely, a game $(A,v)$ is \emph{weakly subadditive} if: $v(C\cup C')\leq v(C)+v(C')$ for any two disjoint coalitions $C,C'$ (i.e., merging any two coalitions is never beneficial). Finally, recall that a game $(A,v)$ is the \emph{sum} of two games, $(A,v_1)$ and $(A,v_2)$, if $v(C)=v_1(C)+v_2(C)$ for all $C\subseteq A$. In this case, we write: $(A,v)=(A,v_2)+(A,v_2)$. Now, we are ready to introduce the following definition.

\begin{bolddefinition} A game $(A,v)$ is (weakly) super-subadditive if it is the sum of two games: a (weakly) superadditive game, denoted by $(A,v^{sup})$, and a (weakly) subadditive game, denoted by $(A,v^{sub})$.
\end{bolddefinition}

The intuition is that $(A,v^{sup})$ represents the rewards from cooperation, which is assumed to increase (weakly) with the size of the coalition. On the other hand, $(A,v^{sub})$ represents the coordination costs, which are also assumed to increase (weakly) with the size of the coalition.


\section{Related Work}\label{sec:relatedWork}

\noindent This section is divided into two subsections. The first describes the $\CFSS$ algorithm of \citet{Bistaffa:etal:14}, while the second describes the $\DyPE$ algorithm of \citet{Vinyals:etal:13}.

\subsection{The CFSS Algorithm}\label{sec:CFSS}
\begin{figure}[th]
\center
\includegraphics[width=14cm]{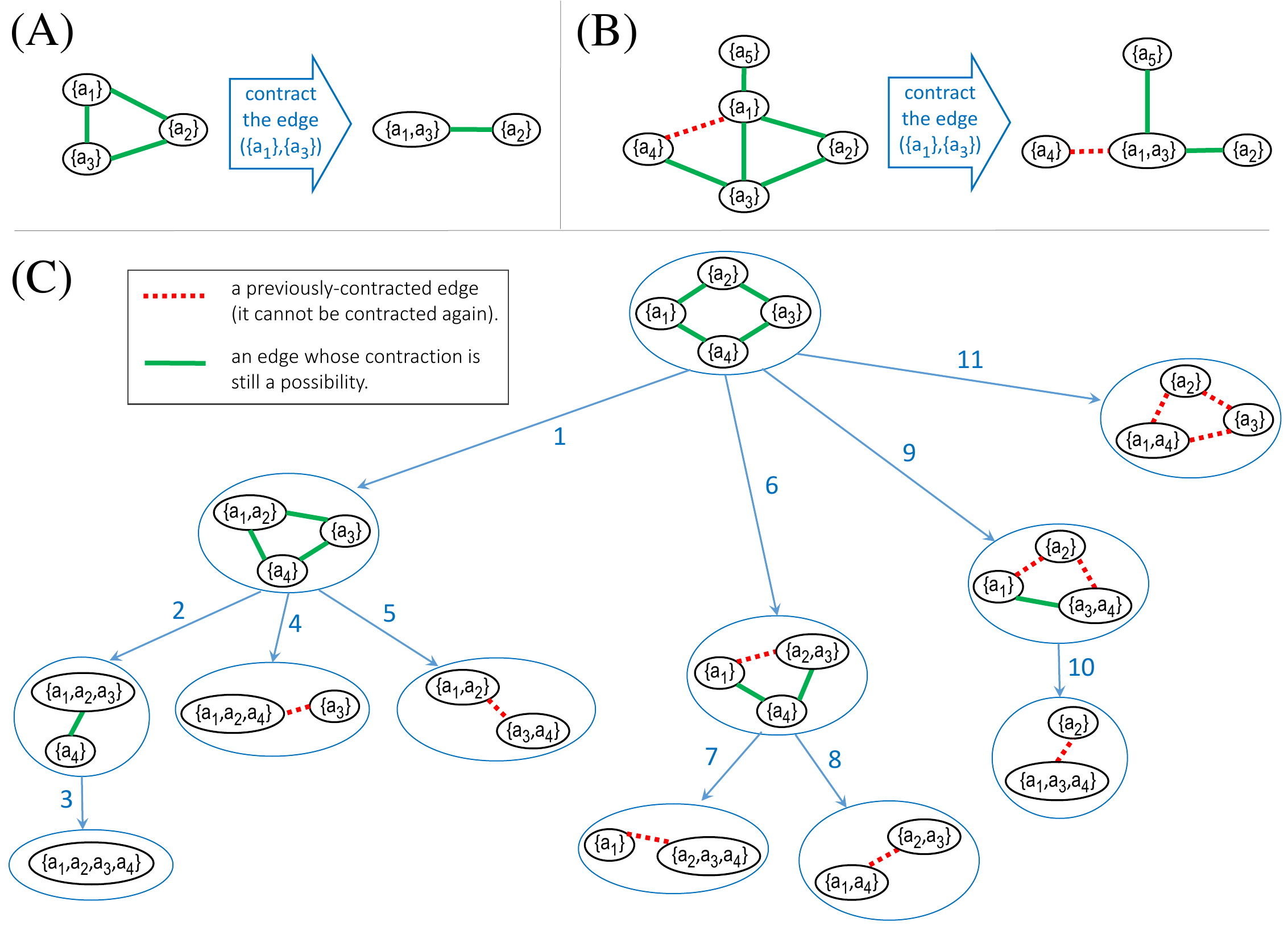}
\caption{Illustration of how Bistaffa et al.'s algorithm works.}
\label{fig:Bistaffa}
\end{figure}

\noindent \citet{Bistaffa:etal:14} proposed the $\CFSS$ algorithm. It is based on \emph{edge contraction}---a basic operation in graph theory which involves: (i) \emph{removing} an edge from a graph, and (ii) \emph{merging} the two nodes that were previously joined by that edge. In our context of graph-restricted games, since every node represents an agent (i.e., a singleton coalition), \emph{``merging the two nodes''} corresponds to \emph{merging the two coalitions} that were represented by those nodes. An example is illustrated in Figure~\ref{fig:Bistaffa}(A).

Taking the entire graph into consideration, the contraction of an edge can be interpreted as a transition from one coalition structure to another. For instance, the contraction of the edge $(\{a_1\},\{a_3\})$ in Figure~\ref{fig:Bistaffa}(B) corresponds to the transition from $\{\{a_1\},\{a_2\},\{a_3\},\{a_4\},\{a_5\}\}$ to $\{\{a_1,a_3\},\{a_2\},\{a_4\},\{a_5\}\}$. Based on this observation, the algorithm repeats the process of contracting different edges, in order to eventually visit all coalition structures. During this process, to ensure that each coalition structure is visited no more than once, the algorithm marks all previously-contracted edges to avoid contracting them again in the future. In Figure~\ref{fig:Bistaffa}, the marked edges are illustrated as dashed lines. Here, it is important to note that the contraction of an edge may result in merging other edges. In Figure~\ref{fig:Bistaffa}(B) for example, contracting $(\{a_1\},\{a_3\})$ results in merging $(\{a_1\},\{a_4\})$ with $(\{a_3\},\{a_4\})$, as well as merging $(\{a_1\},\{a_2\})$ with $(\{a_3\},\{a_2\})$. Whenever this happens, if one of the merged edges happens to be dashed, the edge that results from the merger must also be dashed, again see Figure~\ref{fig:Bistaffa}(B). This ensures that the agents appearing at the two ends of a dashed edge never appear together in the same coalition.

Figure~\ref{fig:Bistaffa}(C) illustrates the sequence in which the algorithm visits all possible coalition structures given the graph $G=(A,E)$ where $A=\{a_1,a_2,a_3,a_4\}$ and $E=\{(a_1,a_2),(a_1,a_4),(a_3,a_2),(a_3,a_4)\}$. Each coalition structure is represented as a node in the illustrated search tree, and the numbers on the edges represent the order in which the algorithm visits the different coalition structures.
Consider the root for example: in its first child we contract $(\{a_1\},\{a_2\})$; in its second child we make $(\{a_1\},\{a_2\})$ dashed, and contract $(\{a_2\},\{a_3\})$; in its third child we make $(\{a_1\},\{a_2\}),(\{a_2\},\{a_3\})$ dashed, and contract $(\{a_3\},\{a_4\})$; finally in its fourth child we make $(\{a_1\},\{a_2\}),(\{a_2\},\{a_3\}),(\{a_3\},\{a_4\})$ dashed, and contract $(\{a_1\},\{a_4\})$.

To speed up the search, a branch-and-bound technique is used whenever the algorithm visits a node---i.e., a partition, $P$---in the search tree. The purpose of this technique is to determine whether it is worthwhile to search $T^{P}$---the sub-tree rooted at $P$. The general idea is to compute an upper bound, denoted $\UB(T^{P})$, on the values of all partitions in $T^{P}$. Then, if this upper bound was not greater than the value of the best partition found so far, then the algorithm avoids searching $T^{P}$. Bistaffa et al. proposed a way of computing $\UB(T^{P})$ for cases where the game under consideration is weakly super-subadditive (see Section~\ref{sec:preliminaries} for more details). In particular, it is possible to compute an upper bound $UB(T^{P})$ based on the following observations:
\begin{itemize} 
\item Every coalition structure in $T^{P}$ is the result of merging some (if not all) of the coalitions in $P$ that are connected via solid edges. Here, the only constraint is that agents appearing at the two ends of a dashed edge must not appear together in the same coalition.
\item Merging coalitions in $P$ can never improve solution quality in a weakly subadditive game. Thus, $V^{sub}(P) = \max_{P\in T^{P}} V^{sub}(P)$.
\item Merging coalitions in $P$ can never reduce solution quality in a weakly superadditive game. Thus, no solution in $T^{P}$ can be better than the solution obtained by: (i) removing all dashed edges, and (ii) merging all coalitions in $P$ that are connected via solid edges. Let us denote this solution as $P^{merge}$. Then, $V^{sup}(P^{merge}) \geq \max_{P\in T^{P}} V^{sup}(P)$.
\end{itemize}
Based on the above observations, we can establish the following upper bound on solution quality: $UB(T^{P})=V^{sub}(P)+V^{sup}(P^{merge})$. This concludes our description of Bistaffa et al.'s algorithm. More details can be found in \cite{Bistaffa:etal:14}.


\subsection{The $\DyPE$ Algorithm}\label{sec:DyPE}

\noindent \citet{Vinyals:etal:13} proposed a dynamic-programming algorithm called $\DyPE$. Before explaining how this algorithm works, let us first briefly describe how dynamic programming works for general games, rather than graph-restricted games. Here is the main idea: to compute an optimal partition of the set of agents, $A$:

\begin{itemize}
\item First, compute an optimal partition of each strict subset of $A$.
\item After that, examine all the possible ways of splitting $A$ into two halves, and replace one of the halves with its optimal partition. More specifically, for every non-empty subset $S\subseteq A:S\neq \emptyset$, split $A$ into two halves, $S$ and $A\setminus S$, and replace $A\setminus S$ with $\opt(A\setminus S)$. Clearly, the union $\{S\}\cup \opt(A\setminus S)$ is a partition of $A$, and the value of this union is $v(C)+v^*(A\setminus S)$. Furthermore, the best such union (i.e., the one with the largest value) is an optimal partition of $A$.
\end{itemize}
Importantly, the above process can be carried out recursively, as captured by the following formula:

\begin{equation}\label{eqn:DP}
v^*(C) = \max_{S\subseteq C: S\neq \emptyset}\big(v(S)+v^*(C\setminus S)\big)
\end{equation}

\begin{figure}[th]
\center
\includegraphics[width=9cm]{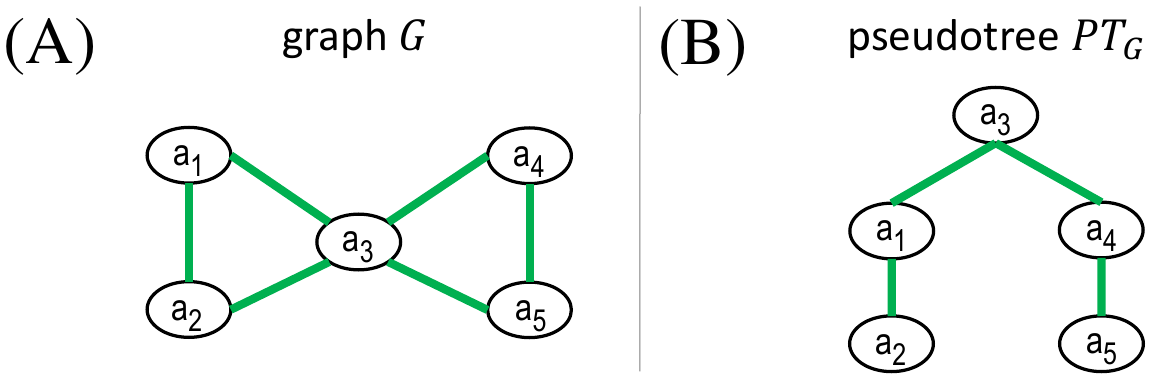}
\caption{A sample graph $G$ and its corresponding pseudotree $\PT_G$.}
\label{fig:Vinyals}
\end{figure}

\noindent Having described a general dynamic programming formula, we now explain how $\DyPE$ speeds up this formula when the game is restricted by a graph. The main idea is to use a \emph{pseudotree}. Basically, given a graph $G=(A,E)$, the pseudotree of $G$, denoted by $\PT_G$, is a rooted tree such that: (i) the set of nodes of $\PT_G$ is the set of agents, and (ii) any two agents who share an edge in $G$ appear on the same branch in $\PT_G$ (an example is illustrated in Figure~\ref{fig:Vinyals}). Let us now explain how $\DyPE$ uses the pseudotree to speed up the formula. To this end, let $b_i$ denote the agent at the $i^{th}$ position of the breadth-first order of nodes in $\PT_G$. In Figure~\ref{fig:Vinyals}(B) for example, that order is: $(a_3,a_1,a_4,a_2,a_5)$, and so $b_1 = a_3$ while $b_4 = a_2$. Now, the broad idea behind $\DyPE$ is to start with the last agent in the breadth-first order, $b_n$, and then move to $b_{n-1}$, then $b_{n-2}$ and so on until it reaches $b_1$. Let $b^{\DyPE}$ denote the agent at which $\DyPE$ has reached in the breadth-first order at any point in time during execution. Then, for each $b^{\DyPE}$, the algorithm solves the following sub-problems:

\begin{equation}\label{eqn:DyPE:subproblems}
\left\{ (C,v,G) \ \big|\ 
  \begin{subarray}{l}
    (C\in \ConnectedSubsets(\{b^{\DyPE},\cdots, b_n\},G)) \\
    \wedge (b^{\DyPE}\in C) \\
    \wedge (A\setminus C\in ConnectedSubsets(A,G))
  \end{subarray}
\right\}
\end{equation}

The pseudo code of $\DyPE$ is shown in Algorithm~\ref{alg:DyPE}. For a proof of the correctness of this algorithm, see \cite{Vinyals:etal:13}.

\RestyleAlgo{ruled}
\begin{algorithm}[thbp]
  \SetAlgoVlined
  \LinesNumbered
  \caption{$\DyPE(A,v,G,\PT_G)$.}
  \label{alg:DyPE}
  \KwIn{A graph-restricted game $(A,v,G)$, and a pseudotree, $\PT_G$.}
  \KwOut{An optimal coalition structure over $A$.}
      \For{$b^{\DyPE}=b_n$ \textbf{to} $b_2$}
      {
          \Comment{iterate over all sub-problems in \eqref{eqn:DyPE:subproblems}:}\\
          \ForEach{$C\in\ConnectedSubsets(\{b^{\DyPE},\cdots, b_n\},G)$ \textbf{such that} $(b^{\DyPE}\in C)\wedge(A\setminus C\in ConnectedSubsets(A,G))$}
          {
              \Comment{Compute $v^*(C)$ and $\mathit{bestSubset}(C)$ (in lines 3 to 8):}\\
              $v^*(C)\gets -\infty$;\\
              \ForEach( \Comment{iterate over all non-empty subsets of $C$ that are each connected in $G$ and contain $b^{\DyPE}$.} ){$S\in\ConnectedSubsets(C,G):b^{\DyPE}\in S$}
              {
                  $\mathit{value}\gets v(S) + \sum_{T\in \mathit{connectedComponents}(C\setminus S)} v^*(T)$; \Comment{Compute the value of $\{S\}\cup\opt(C\setminus S)$, i.e., compute $v(S) + v^*(C\setminus S)$.}\\
                  \If{$v^*(C) < \mathit{value}$}
                  {
                      $v^*(C)\gets \mathit{value}$;\\
                      $\mathit{bestSubset}(C) \gets S$;
                  }
              }
          }
      }
      \Comment{Compute $v^*(A)$ and $\mathit{bestSubset}(A)$ (in lines 9 to 14):}\\
      $v^*(A)\gets -\infty$;\\
      \ForEach( \Comment{iterate over all non-empty subsets of $A$ that are each connected in $G$ and contain $b_1$.} ){$S\in\ConnectedSubsets(A,G):b_1\in S$}
      {
          $\mathit{value}\gets v(S) + \sum_{T\in \mathit{connectedComponents}(A\setminus S)} v^*(T)$;\\
          \If{$v^*(A) < \mathit{value}$}
          {
              $v^*(A)\gets \mathit{value}$;\\
              $\mathit{bestSubset}(A) \gets S$;
          }
      }
      \Comment{Compute an optimal coalition structure over $A$ (in lines 15 to 17):}\\
      $\opt(A)\gets \{A\}$;\\
      \While{$\exists C\in \opt(A): C\neq\mathit{bestSubset}(C)$}
      {
          replace every $C\in \opt(A)$ with $\mathit{bestSubset}(C),C\setminus\mathit{bestSubset}(C)$;
      }
      \Return $\opt(A)$;
\end{algorithm}

\section{Our Tree-Search Algorithm---$\TSP$}\label{sec:TSP}

\noindent As mentioned earlier in the introduction, \citet{Rahwan:etal:12} developed an algorithm for general coalition structure generation problems, which combined a tree-search algorithm with a dynamic-programming algorithm, resulting in a combination that is superior to both its constituent parts. So why not develop a similar hybrid algorithm for graph-restricted games? Perhaps the most natural starting point would be to try and combine $\CFSS$---an existing tree-search algorithm---with $\DyPE$---an existing dynamic-programming algorithm. Unfortunately, however, as we have seen in the above section, both algorithms are based on very different ideas; one is based on \emph{edge contraction}, while the other is based on a \emph{pseudo tree}. As such, the two seem incompatible, or at least hard to combine smoothly. With this in mind, our goal in this section is to develop a tree-search algorithm that can be combined with $\DyPE$. We build our algorithm around the pseudotree representation used by $\DyPE$; the hypothesis here is that if the two algorithms were built around the same representation, it should be possible to combine the two smoothly and effectively. Based on this, we call our algorithm $\TSP$, where TS stands for Tree-Search, and P stands for Pseudotree.

The pseudo code of $\TSP$ can be found in Algorithm~\ref{alg:TSP:main}. In more detail, the algorithm takes as input a graph-restricted game, $(A,v,G)$, and a pseudotree $\PT_G$.  First, in lines 1 to 4, it initializes $\CS^{\dagger}$---the current best solution---to either be equal to $\{A\}$ or $\{\{a_1\},\dots,\{a_n\}\}$, whichever has higher value. After that, in line~5, it uses the parameter $b^{\TSP}$ to iterate over the agents in a breadth-first order in $\PT_G$, starting with $b_2$, and ending with $b_n$.\footnote{\footnotesize{See Section~\ref{sec:DyPE} for more details on the breadth-first order of agents in $\PT_G$.}} Let us denote by $b^{\TSP-1}$ the agents who is just before $b^{\TSP}$ in the breadth-first order. Now, for every $b^{\TSP}$, the algorithm enumerates all the coalitions that are each connected in $G$, and contain every agent in $\{b_1,\cdots, b^{\TSP-1}\}$ but do not contain $b^{\TSP}$ (line~6). For every such coalition, $C$, the algorithm sets the current partition, $P^{\dagger}$, to be equal to $\{C\}$ (line~7). Finally, in line~8, it used the function $\search(\PT^{\dagger}_G, P^{\dagger},\CS^{\dagger})$ to search through the coalition structures that are supersets of $P^{\dagger}$, i.e., the coalition structure that contain $C$. Basically, this recursive function generates different partitions while trying to avoid the unpromising ones using a branch-and-bound technique. Next, we explain how this function works.

\RestyleAlgo{ruled}
\begin{algorithm}[thbp]
  \SetAlgoVlined
  \LinesNumbered
  \caption{$\TSP(A,v,G,\PT_G)$.}
  \label{alg:TSP:main}
  \KwIn{A graph-restricted game $(A,v,G)$, and a pseudotree $\PT_G$.}
  \KwOut{An optimal coalition structure over $A$.}
      \Comment{initialize $\CS^{\dagger}$---the current best solution (lines 1 to 4).}\\
      \If{$V(\{A\}) > V(\{\{a_1\},\dots,\{a_n\}\})$}
      {
          $\CS^{\dagger}\gets \{A\}$;
      }
      \Else
      {
            $\CS^{\dagger}\gets \{\{a_1\},\dots\{a_n\}\}$;
      }
      \Comment{Search through different coalition structures (lines 5 to 8).}\\
      \For{$b^{\TSP}=b_2$ \textbf{to} $b_n$}
      {
          \Comment{iterate over all non-empty subsets of $A$ that are each connected in $G$ and do not contain $b^{\TSP}$ but contain every agent before $b^{\TSP}$ in the breadth-first order.}\\
          \ForEach{$C\in\ConnectedSubsets(A,G):\{b_1,\cdots, b^{\TSP-1}\}\subseteq C\subseteq A\setminus\{b^{\TSP}\}$}
          {
              $P^{\dagger}\gets \{C\}$; \Comment{initialize $P^{\dagger}$---the current partition.}\\
              $\CS^{\dagger} \gets \search(\PT^{\dagger}_G,P^{\dagger},\CS^{\dagger})$; \Comment{updated $\CS^{\dagger}$ by searching through the partitions of $A$ that are supersets of $P^{\dagger}$.}\\
          }
      }
      \Return $\CS^{\dagger}$;
\end{algorithm}

The pseudo code of $\search(\PT^{\dagger}_G, P^{\dagger},\CS^{\dagger})$ is given in Algorithm~\ref{alg:TSP:search}. Here, $C^{\dagger}$ denotes the agents that are not in $P^{\dagger}$ (line~1 of Algorithm~\ref{alg:TSP:search}), while $a^{\dagger}$ denotes the first agent in the breadth-first order who is not in $P^{\dagger}$ (line~2). Then, out of all the connected coalitions that can be added to $P^{\dagger}$, the algorithm always starts by adding to $P^{\dagger}$ a coalition containing $a^{\dagger}$ (lines 3 and 4). Now if the new $P^{\dagger}$ is a coalition structure over $A$, then the algorithm updates $\CS^{\dagger}$---the best solution found so far (lines 5 to 7). Otherwise, it computes an upper bound on the value of every partition of $A$ that is a superset of the new $P^{\dagger}$ (line~9). Based on this upper bound, the algorithm determines whether it is worthwhile to consider adding more coalitions to the new $P^{\dagger}$ (line~10). If so, then it makes a recursive call with the new $P^{\dagger}$ (line~{11}). The function $\mathbf{computeUpperBound}(P^{\dagger})$---which computes the aforementioned upper bound---can be specified based on any additional domain knowledge. For instance, if the game is known to be super-subadditive, then this function may return: $V(P^{\dagger}) + v^{sup}(C^{\dagger}) + \sum_{a_i\in C^{\dagger}}v^{sub}(\{a_i\})$.

\RestyleAlgo{ruled}
\begin{algorithm}[t]
  \SetAlgoVlined
  \LinesNumbered
  \caption{$\search(\PT_G, P^{\dagger}, \CS^{\dagger})$---a function used in $\TSP$.}
  \label{alg:TSP:search}
  \KwIn{$\PT_G$---the pseudotree, $P^{\dagger}$---the current partition (which does not yet contain all agents in $A$), and $\CS^{\dagger}$---the current best solution.}
  \KwOut{The best partition of $A$ that is a superset of $P^{\dagger}$.}
      $C^{\dagger} \gets A\setminus \bigcup P^{\dagger}$;\Comment{i.e., $C^{\dagger}$ consists of all agents not in $P^{\dagger}$.}\\
      $a^{\dagger} \gets b_i: (b_i\in C^{\dagger})\wedge(\{b_1,\cdots,b_{i-1}\}\cap C^{\dagger}=\emptyset)$;\Comment{i.e., $a^{\dagger}$ is the first agent in the breadth-first order who is not in $P^{\dagger}$.}\\
      \ForEach( \Comment{iterate over all non-empty subsets of $C^{\dagger}$ that are each connected in $G$ and contain $a^{\dagger}$.} ){$C\in\ConnectedSubsets(C^{\dagger},G):a^{\dagger}\in C$}
      {
          $P^{\dagger} \gets P^{\dagger}\cup \{C\};$ $C^{\dagger} \gets A\setminus C$; \Comment{add $C$ to $P^{\dagger}$ and remove it from $C^{\dagger}$.}\\
          \If( \Comment{if $P^{\dagger}$ is a coalition structure over $A$.} ){$C^{\dagger} = \emptyset$}
          {
              \If{$V(\CS^{\dagger}) < V(P^{\dagger})$}
              {
                  $\CS^{\dagger} \gets P^{\dagger};$\\
              }
          }
          \Else
          {
              $\UB \gets \mathbf{computeUpperBound}(P^{\dagger})$ \Comment{compute an upper bound on the value of every partition of $A$ that is a superset of $P^{\dagger}$}\\
              \If( \Comment{apply the branch-and-bound technique.} ){ $V(\CS^{\dagger}) < \UB$ }
              {
                  $\search(\PT^{\dagger}_G,P^{\dagger},\CS^{\dagger});$ \Comment{recursive call.}\\
              }
          }
          $P^{\dagger} \gets P^{\dagger}\setminus \{C\};$ $C^{\dagger} \gets A\cup C$; \Comment{remove $C$ from $P^{\dagger}$ and add it to $C^{\dagger}$.}\\
      }
      \Return $\CS^{\dagger}$;
\end{algorithm}

\begin{figure}[th]
\center
\includegraphics[width=12cm]{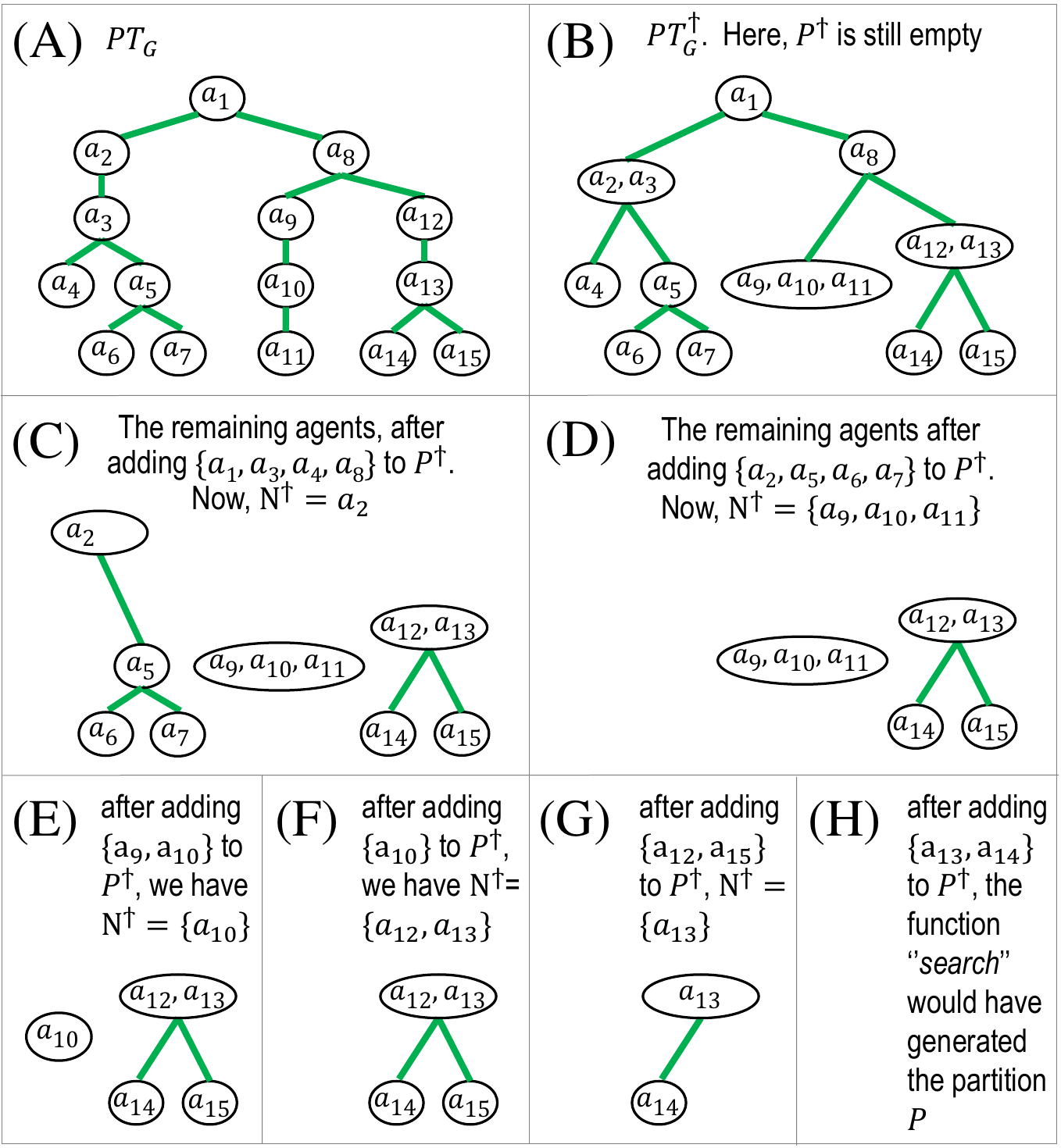}
\caption{The illustration.}
\label{fig:TSP}
\end{figure}

\section{Our Hybrid Algorithm---$\DTSP$}\label{sec:D-TSP}

\noindent In this section, we present $\DTSP$---a hybrid algorithm that combines $\DyPE$ with $\TSP$ in a way that obtains the best features of both. First, we introduce the necessary modifications of each algorithm (Subsections \ref{sec:DyPE*} and \ref{sec:TSP*}), and then show how to combine the modified versions (Subsection~\ref{sec:DyPEsubsection}).

\subsection{$\DyPE^*$---a Modified Version of $\DyPE$}\label{sec:DyPE*}

\noindent In this subsection, we modify $\DyPE$ such that it becomes an anytime algorihtm, i.e., it does not only return a solution \emph{after} termination, but also returns interim solutions \emph{during} execution. This clearly adds more resilience against failure. For example, if the algorithm runs out of memory during execution, then instead of wasting all the effort that the algorithm has put before the failure, it would at least return a valid solution using all the sub-problems that it has already solved.

Before introducing our modifications, let us first revisit $\DyPE$ and analyze the way it works. Looking at Algorithm~\ref{alg:DyPE}, one can see that $\DyPE$ ultimately boils down to the following main steps:

\begin{itemize}
\item \textbf{Step 1:} with $b^{\DyPE}$ running from $b_n$ to $b_2$, solve the following sub-problems: $(C,v,G)$ such that $C\subseteq\{b^{\DyPE},\dots,b_n\}$ and $b^{\DyPE}\in C$ and $A\setminus C$ is connected (lines 1 to 8 of Algorithm~\ref{alg:DyPE}).
\item \textbf{Step 2:} for each subset $S\in\ConnectedSubsets(A,G):b_1\in S$, compute the value of the best coalition structure containing $S$ (lines 10 and 11 of Algorithm~\ref{alg:DyPE}).
\end{itemize}

The problem with the above process is that $\DyPE$ does not examine a single coalition structure over $A$ until it has finished Step~1---a step which involves solving sub-problems the number of which may be exponential (depending on the topology of the graph). Let us now consider a sample subset that the algorithm encounters during Step~2, given a problem of $7$ agents. Let this subset be $S=\{b_1,b_2,b_3,b_6\}$. When the algorithm encounters this particular $S$, it will compute the value of the best coalition structure containing $\{b_1,b_2,b_3,b_6\}$, using the already-computed solutions to the following sub-problems: $(T,v,G)$ where $T$ is a connected component in the sub-graph induced by $\{b_4,b_5,b_7\}$ (see line~11 of Algorithm~\ref{alg:DyPE}). Our critical observation is that the solutions to the aforementioned sub-problems were all computed when $\DyPE$ finished dealing with $b^{\DyPE}=b_4$ in Step~1. More specifically, at that moment, $\DyPE$ has already solved the following sub-problems:
\begin{itemize}
\item $(C,v,G)$ where $A\setminus C$ is connected and $b_7\in C$ and $C\subseteq\{b_7\}$;
\item $(C,v,G)$ where $A\setminus C$ is connected and $b_6\in C$ and $C\subseteq\{b_6,b_7\}$;
\item $(C,v,G)$ where $A\setminus C$ is connected and $b_5\in C$ and $C\subseteq\{b_5,b_6,b_7\}$;
\item $(C,v,G)$ where $A\setminus C$ is connected and $b_4\in C$ and $C\subseteq\{b_4,b_5,b_6,b_7\}$.
\end{itemize}
The above sub-problems surely include every $(T,v,G)$ where $T$ is a connected component in the sub-graph induced by $\{b_4,b_5,b_7\}$. In other words, after solving the above sub-problems, $\DyPE$ had all the information needed to compute the value of the best coalition structure containing $\{b_1,b_2,b_3,b_6\}$. This suggests that $\DyPE$ can be modified such that it examines certain coalition structures \emph{during} Step~1, not \emph{after} Step~1.

Based on the above observation, we modify $\DyPE$ such that, instead of following the above two steps, it follows txhis one:
\begin{itemize}
\item \textbf{Step 1:} with $b^{\DyPE}$ running from $b_n$ to $b_2$:
    \begin{itemize}
    \item \textbf{Step 1.1:} solve the following sub-problems: $(C,v,G)$ such that $C\subseteq\{b^{\DyPE},\dots,b_n\}$ and $b^{\DyPE}\in C$ and $A\setminus C$ is connected.
    \item \textbf{Step 1.2:} for each subset $S\in\ConnectedSubsets(A,G)$ such that $b^{\DyPE}\notin S$ and $\{b_1,\dots, b^{\DyPE-1}\}\subseteq S$, compute the value of the best coalition structure containing $S$.
    \end{itemize}
\end{itemize}
One can easily see that, with the above steps, the algorithm will never consider the same $S$ more than once. Moreover, whenever a certain $S$ is encountered, all relevant sub-problems of $A\setminus S$ have already been solved, including every $(T,v,G)$ where $T$ is a connected component in the sub-graph induced by $A\setminus S$.

We call the modified version $\DyPE^*$. The pseudo code is provided in Algorithm~\ref{alg:DyPE*}. As can be seen, this an anytime algorithm, unlike $\DyPE$.
 
\RestyleAlgo{ruled}
\begin{algorithm}[thbp]
  \SetAlgoVlined
  \LinesNumbered
  \caption{$\DyPE^*(A,v,G,\PT_G)$.}
  \label{alg:DyPE*}
  \KwIn{A graph-restricted game $(A,v,G)$, and a pseudotree, $\PT_G$.}
  \KwOut{An optimal coalition structure over $A$.}
      $\CS^{\dagger}\gets\{A\}$;\Comment{initialize $\CS^{\dagger}$---the current best solution, which is needed in this anytime version of DyPE.}\\
      \For{$b^{\DyPE}=b_n$ \textbf{to} $b_2$}
      {
          \Comment{iterate over all sub-problems in \eqref{eqn:DyPE:subproblems}:}\\
          \ForEach{$C\in\ConnectedSubsets(\{b^{\DyPE},\cdots, b_n\},G)$ \textbf{such that} $(b^{\DyPE}\in C)\wedge(A\setminus C\in ConnectedSubsets(A,G))$}
          {
              \Comment{Compute $v^*(C)$ and $\mathit{bestSubset}(C)$ (in lines 4 to 8):}\\
              $v^*(C)\gets -\infty$;\\
              \ForEach( \Comment{iterate over all non-empty subsets of $C$ that are each connected in $G$ and contain $b^{\DyPE}$.} ){$S\in\ConnectedSubsets(C,G):b^{\DyPE}\in S$}
              {
                  $\mathit{value}\gets v(S) + \sum_{T\in \mathit{connectedComponents}(C\setminus S)} v^*(T)$;\Comment{Compute the value of $\{S\}\cup\opt(C\setminus S)$, i.e., compute $v(S) + v^*(C\setminus S)$.}\\
                  \If{$v^*(C) < \mathit{value}$}
                  {
                      $v^*(C)\gets \mathit{value}$;\\
                      $\mathit{bestSubset}(C) \gets S$;
                  }
              }
          }
          \Comment{Search every coalition structure containing a connected coalition whose members include $b_1,\cdots, b^{\DyPE-1}$, but not $b^{\DyPE}$:}\\         
          \ForEach{$S\in\ConnectedSubsets(A,G): \{b_1,\cdots,b^{\DyPE-1}\}\subseteq S \subseteq A\setminus\{b^{\DyPE}\}$}
          {
              $\mathit{value}\gets v(S) + \sum_{T\in \mathit{connectedComponents}(A\setminus S)} v^*(T)$;\\
              \If{$v^*(A) < \mathit{value}$}
              {
                  $v^*(A)\gets \mathit{value}$;\\
                  $\mathit{bestSubset}(A) \gets S$;\\
                  $\CS^{\dagger}\gets \{S\} \cup \mathit{connectedComponents}(A\setminus S)$;\\
                  \While{$\exists C\in \CS^{\dagger}: C\neq\mathit{bestSubset}(C)$}
                  {
                      replace every $C\in \CS^{\dagger}$ with $\mathit{bestSubset}(C)$ and $C\setminus\mathit{bestSubset}(C)$;
                  }
              }
          }
      }
      \Return $\CS^{\dagger}$;
\end{algorithm}

\subsection{$\TSP^*$---a Modified Version of $\TSP$}\label{sec:TSP*}

\noindent Our goal in this subsection is to modify $\TSP$ such that it can take advantage of any solutions to sub-problems that were already computed by $\DyPE^*$. To this end, let us first analyze how $\TSP$ works. Looking at Algorithm~\ref{alg:TSP:main}, one can see that $\TSP$ ultimately boils down to the following main steps:
\begin{itemize}
\item\textbf{Step~1:} with $b^{\TSP}$ running from $b_2$ to $b_n$, set the \emph{current partition} $P^{\dagger}$ to be equal to some $\{C\}$, where $C$ is a connected coalition that does not contain $b^{\TSP}$, but contains all of: $b_1,\dots, b^{\TSP-1}$ (lines 5 to 7 of Algorithm~\ref{alg:TSP:main}).\footnote{\footnotesize Recall that $b^{\TSP-1}$ denotes the agent just before $b^{\TSP}$ in the breadth-first order of agents in the pseudo tree $\PT^G$.}
    \begin{itemize}
    \item\textbf{Step~1.1:} keep adding different coalitions to $P^{\dagger}$, thus obtaining different coalition structures over $A$ (line~8 of Algorithm~\ref{alg:TSP:main}). Any coalition added to $P^{\dagger}$ must contain $a^{\dagger}$---the first agent in the breadth-first order who is not already in $P^{\dagger}$ (see line~2 of Algorithm~\ref{alg:TSP:search}). Every time a new coalition is added to $P^{\dagger}$, a branch-and-bound technique is used to check whether the coalitions that are in $P^{\dagger}$ are promising (lines 9 to 11 of Algorithm~\ref{alg:TSP:search}).
    \end{itemize}
\end{itemize}
During the above process, for any given $P^{\dagger}$, the algorithm will try all possible coalition structures that are supersets of $P^{\dagger}$, except those that are deemed unpromising by the branch-and-bound technique. In other words, it will try adding to $P^{\dagger}$ every promising partition of $A\setminus \bigcup P^{\dagger}$. Importantly, however, if we were to run $\DyPE^*$ in parallel with $\TSP$, then the latter algorithm may be able to construct an optimal partition of $A\setminus \bigcup P^{\dagger}$ easily using the partial results of the former. This is based on the following two observations:
\begin{itemize}
\item $A\setminus \bigcup P^{\dagger}\subseteq\{a^{\dagger},\dots, b_n\}$. This is simply because $a^{\dagger}$ is by definition the first agent in the breadth-first order who is not in $P^{\dagger}$.
\item if the current $b^{\DyPE}$ happens to be before $a^{\dagger}$ in the breadth-first order, then $\DyPE^*$ has already computed all relevant sub-problems $(C,v,G)$ such that $C\subseteq \{a^{\dagger},\dots, b_n\}$ (see Sectino~\ref{sec:DyPE*} for more details). 
\end{itemize}
Based on the above observations, we propose a modified version of $\TSP$, called $\TSP^*$, which works as follows. Whenever $b^{\DyPE}$ happens to be before $a^{\dagger}$ in the breadth-first order, $\TSP^*$ does not try the different partitions of $A\setminus\bigcup P^{\dagger}$, but instead computes the value of an optimal such partition as follows:
$$
V^*(A\setminus \bigcup P^{\dagger}) = \sum_{T\in \mathit{connectedComponents}(A\setminus \bigcup P)} v^*(T).
$$
Now, if $V(P)+V^*(A\setminus \bigcup P^{\dagger})$ happens to be greater than $V(\CS^{\dagger})$---the value of the current best solution, then $\TSP^*$ needs to compute a coalition structure $P^{\dagger}\cup \opt(A\setminus\bigcup P^{\dagger})$ because it is better than $\CS^{\dagger}$. This computation can be done as follows. First, the algorithm sets $\CS^{\dagger}$ to be equal to $P^{\dagger}\cup \mathit{connectedComponents}(A\setminus \bigcup P)$, and then iteratively replaces every $C\in\CS^{\dagger}$ with $\mathit{bestSubset}(C)$ and $A\setminus\mathit{bestSubset}(C)$. This is done until $C=\mathit{bestSubset}(C)$ for all $C\in\CS^{\dagger}$.

\subsection{Combining $\DyPE^*$ with $\TSP^*$}\label{sec:DyPEsubsection}

\noindent In this subsection, we introduce $\DTSP$, an algorithm that runs both $\DyPE^*$ and $\TSP^*$ in parallel, such that they aid each other during the search. Basically, $\DTSP$ is based on the following observations:
\begin{itemize}
\item $\DyPE^*$ solves sub-problems in the following sequence (see Sectino~\ref{sec:DyPE*} for more details). With $b^{\DyPE}$ running from $b_n$ to $b_2$
    \begin{itemize}
    \item it solves the sub-problems: $(C,v,G)$ such that $C\subseteq\{b^{\DyPE},\dots,b_n\}$ and $b^{\DyPE}\in C$ and $A\setminus C$ is connected (see lines 2 and 3 of Algorithm~\ref{alg:DyPE*}).
    \item it searches all coalition structures containing a connected coalition $C$ where:
    $$\{b_1,\dots, b^{\DyPE-1}\}\subseteq C\subseteq A\setminus \{b^{\DyPE}\}.$$
    \end{itemize}
\item We deliberately designed $\TSP^*$ such that it searches coalition structures in the following sequence. With $b^{\TSP}$ running from $b_2$ to $b_n$, it searches all coalition structures containing a connected coalition $C$ where:
$$\{b_1,\cdots, b^{\TSP-1}\}\subseteq C \subseteq A\setminus\{b^{\TSP}\}$$.
\end{itemize}

Note that $b^{\DyPE}$ runs from $b_n$ to $b_2$, while $b^{\TSP}$ runs from $b_2$ to $b_n$. Thus, based on the above observations, when the position of $b^{\DyPE}$ becomes smaller than that of $b^{\TSP}$, the algorithms $\DyPE^*$ and $\TSP^*$ would have jointly searched the entire space, at which case $\DTSP$ terminates.


\section{Performance Evaluation}\label{sec:evaluation}

\begin{figure}[th]
\center
\includegraphics[width=12cm]{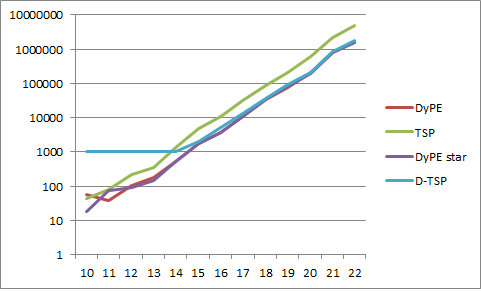}
\caption{Very preliminary simulation results. The number of agents runs from 10 to 22 ($x$-axis.}
\label{fig:results}
\end{figure}

Figure~\ref{fig:results} presents very preliminary simulation results.



\section{Conclusions and Future Work}\label{sec:conclusions}

\noindent Our aim was to develop a coalition structure generation problem for graph-restricted games. Our inspiration came from an algorithm for general coalition structure generation problems, which combined a dynamic-programming algorithm with a tree-search algorithm, resulting in a combination that is superior to both its constituent parts \cite{Rahwan:etal:12}. Following these guidelines, we developed a tree-search algorithm, called $\TSP$ to be compatible with an existing dynamic-programming algorithm, called $\DyPE$ \cite{Vinyals:etal:13}. After that, we showed how to modify the two algorithms such that they are compatible with each other. Specifically, we modified $\DyPE$ to make an anytime algorithm that returns interim solutions, and modified $\TSP$ such that it solutions to sub-problems that were computed by $\DyPE$ at any point in time. After that, we showed that the modified version of $\DyPE$ gradually covers the search space from a certain direction, while the modified version of $\TSP$ gradually covers the search space from the opposite direction; the two algorithms terminate when they meet each other somewhere in the middle. This way, the portion searched by each algorithm will naturally reflect its relative strength on the problem instance at hand. Our future work involves evaluating $\DTSP$ empirically on a wider range of graph-restricted games.


\section*{Acknowledgements}

\noindent This work was supported by the Polish National Science Centre grant number
2014/13/B/ST6/01807. Tomasz P. Michalak was also supported
by the European Research Council under Advanced
Grant 291528 (``RACE'').

{\fontsize{10}{10}\selectfont{

\begin{thebibliography}{}

\bibitem[\protect\citeauthoryear{Bistaffa, Farinelli, Cerquides,
  Rodr\'{\i}guez-Aguilar, and Ramchurn}{Bistaffa
  et~al\mbox{.}}{2014}]{Bistaffa:etal:14}
{\sc Bistaffa, F.}, {\sc Farinelli, A.}, {\sc Cerquides, J.}, {\sc
  Rodr\'{\i}guez-Aguilar, J.}, {\sc and} {\sc Ramchurn, S.~D.} 2014.
\newblock Anytime coalition structure generation on synergy graphs.
\newblock In {\em Proceedings of the 13th International Conference on
  Autonomous Agents and Multi-agent Systems}. AAMAS '14. 13--20.

\bibitem[\protect\citeauthoryear{Michalak, Rahwan, Sroka, Dowell, Wooldridge,
  McBurney, and Jennings}{Michalak et~al\mbox{.}}{2009}]{Michalak:09}
{\sc Michalak, T.}, {\sc Rahwan, T.}, {\sc Sroka, J.}, {\sc Dowell, A.}, {\sc
  Wooldridge, M.}, {\sc McBurney, P.}, {\sc and} {\sc Jennings, N.~R.} 2009.
\newblock On representing coalitional games with externalities.
\newblock In {\em ACM EC '09: Tenth ACM Conference on Electronic Commerce}.
  11--20.

\bibitem[\protect\citeauthoryear{Myerson}{Myerson}{1977}]{Myerson:1977}
{\sc Myerson, R.} 1977.
\newblock Graphs and cooperation in games.
\newblock {\em Mathematics of Operations Research\/}~{\em 2,\/}~3, 225--229.

\bibitem[\protect\citeauthoryear{Rahwan and Jennings}{Rahwan and
  Jennings}{2008}]{Rahwan:Jennings:08b}
{\sc Rahwan, T.} {\sc and} {\sc Jennings, N.~R.} 2008.
\newblock An improved dynamic programming algorithm for coalition structure
  generation.
\newblock In {\em {AAMAS'08: Seventh International Conference on Autonomous
  Agents and Multi-Agent Systems}}. 1417--1420.

\bibitem[\protect\citeauthoryear{Rahwan, Michalak, Elkind, Wooldridge, and
  Jennings}{Rahwan et~al\mbox{.}}{2014}]{Rahwan:etal:14}
{\sc Rahwan, T.}, {\sc Michalak, T.}, {\sc Elkind, E.}, {\sc Wooldridge, M.},
  {\sc and} {\sc Jennings, N.~R.} 2014.
\newblock An exact algorithm for coalition structure generation and complete
  set partitioning.
\newblock {\em
  http://www.cs.ox.ac.uk/publications/publication6962-abstract.html\/}.

\bibitem[\protect\citeauthoryear{Rahwan, Michalak, and Jennings}{Rahwan
  et~al\mbox{.}}{2012}]{Rahwan:etal:12}
{\sc Rahwan, T.}, {\sc Michalak, T.}, {\sc and} {\sc Jennings, N.~R.} 2012.
\newblock A hybrid algorithm for coalition structure generation.
\newblock In {\em Proceedings of the 26th AAAI Conference on Artificial
  Intelligence (AAAI-2012)}.

\bibitem[\protect\citeauthoryear{Rahwan, Michalak, Wooldridge, and
  Jennings}{Rahwan et~al\mbox{.}}{2012}]{Rahwan:etal:12:AIJ}
{\sc Rahwan, T.}, {\sc Michalak, T.}, {\sc Wooldridge, M.}, {\sc and} {\sc
  Jennings, N.~R.} 2012.
\newblock Anytime coalition structure generation in multi-agent systems with
  positive or negative externalities.
\newblock {\em Artificial Intelligence\/}~{\em 186,\/}~0, 95 -- 122.

\bibitem[\protect\citeauthoryear{Rahwan, Michalak, Wooldridge, and
  Jennings}{Rahwan et~al\mbox{.}}{2016}]{Rahwan:etal:15:AIJ}
{\sc Rahwan, T.}, {\sc Michalak, T.}, {\sc Wooldridge, M.}, {\sc and} {\sc
  Jennings, N.~R.} 2015.
\newblock Coalition structure generation: A survey.
\newblock {\em Artificial Intelligence\/}~{\em 229,\/}~0, 139 -- 174.


\bibitem[\protect\citeauthoryear{Rahwan, Ramchurn, Giovannucci, and
  Jennings}{Rahwan et~al\mbox{.}}{2009}]{Rahwan:09a}
{\sc Rahwan, T.}, {\sc Ramchurn, S.~D.}, {\sc Giovannucci, A.}, {\sc and} {\sc
  Jennings, N.~R.} 2009.
\newblock An anytime algorithm for optimal coalition structure generation.
\newblock {\em Journal of Artificial Intelligence Research (JAIR)\/}~{\em 34},
  521--567.

\bibitem[\protect\citeauthoryear{Vinyals, Voice, Ramchurn, and
  Jennings}{Vinyals et~al\mbox{.}}{2013}]{Vinyals:etal:13}
{\sc Vinyals, M.}, {\sc Voice, T.}, {\sc Ramchurn, S.}, {\sc and} {\sc
  Jennings, N.~R.} 2013.
\newblock A hierarchical dynamic programming algorithm for optimal coalition
  structure generation.
\newblock {\em http://arxiv.org/abs/1310.6704\/}.

\end{thebibliography}

}}

\end{document}